# High Temperature Phase Stability in $Li_{0.12}Na_{0.88}NbO_3$: A Combined powder X-Ray and Neutron Diffraction Study

S. K. Mishra, P. S. R. Krishna, A. B. Shinde, V. B. Jayakrishnan, R. Mittal, P. U. Sastry and S. L. Chaplot

*Solid State Physics Division, Bhabha Atomic Research Centre, Trombay, Mumbai 400 085, India.*

## Abstract

Lithium doped sodium niobate is an ecofriendly piezoelectric material that exhibits a variety of structural phase transitions with composition and temperature. We have investigated the phase stabilities of an important composition $Li_{0.12}Na_{0.88}NbO_3$ (LNN12) using a combination of powder x-ray and neutron diffraction techniques in the temperature range 300 - 1100 K. Detailed Rietveld analyses of thermo-diffractograms show a variety of structural phase transitions ranging from non-polar antiferrodistortive to ferroelectric in nature. In the temperature range of 525 K to 675 K, unambiguous experimental evidence is shown for phase coexistence of orthorhombic paraelectric O1 phase (space group ***Cmcm***) and orthorhombic ferroelectric O2 phase (space group ***Pmc2$_1$***). The $b_p$ primitive lattice parameter of the ferroelectric orthorhombic phase (O2 phase) decreases, while the $a_p$ and $c_p$ primitive lattice parameters show normal increase with increase in temperature. Above 675 K, in the O1 phase, all lattice parameters come close to each other and increase continuously with increase of temperature, and around 925 K, $a_p$ parameter approaches $b_p$ parameter and thus the sample undergoes an orthorhombic to tetragonal phase transition. Further as temperature increases, the $c_p$ lattice parameter decreases, and finally approaches to $a_p$ parameter, and the sample transform into the cubic phase. The continuous change in the lattice parameters reveals that the successive phase transformations from orthorhombic O1 to high temperature tetragonal phase and finally to the cubic phase are not of a strong first order type in nature. We argue that application of chemical pressure as a result of Li substitution in $NaNbO_3$ matrix, favours the freezing of zone centre phonons over the zone boundary phonons that are known to freeze in pure $NaNbO_3$ as function of temperature.

# 1. Introduction

It is extremely important to find ecofriendly materials to satisfy the ever-growing energy demands. Piezoelectric materials transform ambient vibrations or movement into electrical energy and can be used to power up micro scale devices like: mobile phone, sensors/actuators, telemetry and MEMS. As a result of environmental regulations, it is needed to develop lead-free piezoelectric materials. For this purpose, solid solution of alkaline niobates ($K_xNa_{1-x}NbO_3$, $Li_xNa_{1-x}NbO_3$) are suitable candidates because they possess comparable morphotropic phase boundary and have ultra-large piezoresponse similar to Pb based materials [1-4].

Solid solution of lithium sodium niobate ($Li_xNa_{1-x}NbO_3$ referred after as LNNx) have been extensively investigated using thermal expansion [5], microstructure [6], dielectric properties [7-9], electromechanical properties [7,10], pyroelectric properties [11], low temperature properties [12], high-temperature calorimetry [13] and the X-ray and neutron powder diffraction techniques [14-17]. Diffraction studies of LNNx showed that it undergoes a series of structural phase transitions ranging from ferroelectric to antiferroelectric, antiferroelectric to paraelectric, ferroelectric to paraelectric and ferroelastic- paraelectric as a function of composition(x) and temperatures [7-17]. At ambient conditions, Krainik *et. al.* [6] have shown that LNNx exhibits antiferroelectric phase up to x=0.01, and by further increasing concentration of Li further, the ferroelectric phase is stabilized. Nitta [16] reported a structural transition from pseudo-monoclinic to pseudo-tetragonal around x=0.12 (LNN12) at room-temperature. Henson *et. al.* [10] observed anomalies in dielectric permitivity and electromechanical coupling coefficient at around x=0.12 at room temperature. They proposed that observed anomaly was attributed to the transition from orthorhombic to tetragonal phase. Y.D. Juang *et. al.* [17] performed Raman and X-ray diffraction (XRD) measurements and they did not find any signature for presence of the morphotropic phase boundary (between orthorhombic and tetragonal) or rhombohedral phase at x=0.12 from their Raman scattering and x-ray diffraction (XRD) measurements. However, Yuzyuk *et. al.* [4] re-examined the LNNx ceramics using synchrotron X-ray and Raman scattering and reported existence of rhombohedral phase in a narrow concentration interval around x=0.12 at room temperature. More recently, Peel *et. al.* [15] studied LNNx by combined X-ray and neutron powder diffraction as well as by $^{23}$Na solid-state NMR spectroscopy. They reported a single polar orthorhombic phase for x = 0.05 and phase coexistence (orthorhombic and rhombohedral phase) for $0.08 \leq x \leq 0.20$. They also showed that the relative phase fraction is dependent on synthesis conditions.

Beyond the technological applications, alkaline niobates have been a rich model system for understanding of the mechanism of structural phase transitions. For example one of the end members, sodium niobate exhibits a most complex sequence of structural phase transitions in the perovskite family [18-19]. Using detailed and systematic temperature dependent neutron diffraction studies (T= 17 to 1050

K), we showed that above 913 K, it has a paraelectric cubic phase (***Pm3m***) [17-18]. On lowering the temperature it undergoes structural phase transition to a series of antiferrodistortive phases in following order: tetragonal (*T*2) ***P4/mbm***, orthorhombic (*T*1) ***Cmcm***, orthorhombic (*S*) ***Pbnm***, orthorhombic (*R*) ***Pbnm***, orthorhombic (*P*) ***Pbcm*** phases, and a rhombohedral ***R3c*** phase [17-18]. We presented unambiguous evidence for the coexistence of a ferroelectric (FE) ***R3c*** phase and an antiferroelectric (AFE) phase (***Pbcm***) over a wide range of temperatures. We also carried out high pressure neutron diffraction measurements up to 11 GPa at ambient temperature which shows a transition from the ***Pbcm*** to the ***Pbnm*** phase [20]. These transitions are characterized by the appearance and the disappearance of superlattice reflections in the powder diffraction patterns. These superlattice reflections originate from the condensation of zone-centre and zone-boundary phonon modes.

First principles calculation on these materials reveals that the structural phase transitions originate from the competing interactions between different phonon instabilities occurring in the cubic phase [21]. The cubic perovskite structure is unstable to energy-lowering distortions like zone-center distortions (resulting in ferroelectricity), zone-boundary distortions involving rotations and/or tilting of the oxygen octahedral. Depending on thermodynamic conditions various phases can be stabilized. For example application of chemical pressure (doping of Li/K in $NaNbO_3$) promotes zone center (i.e. ferroelectricity) and destabilizes the zone boundary distortion (multiplicity of cell are reduced).

Despite these extensive experimental studies as a function of composition, there are numerous controversies surrounding the phase diagram of Li modified $NaNbO_3$ especially, on the existence of rhombohedral phase. Recently LNN12 has got considerable attention due to its ferroelectric phase coexistence which is termed similar to an existence of morphotrophic phase boundary as observed in PZT. $Li_{0.12}Na_{0.88}NbO_3$ (LNN12) sample was prepared by different methods and characterized by different techniques but their main focus was limited to its electrical properties only. Since properties of the materials are strongly dependent on their structure, it is desirable to investigate the structural phase stabilities as a function of temperature. Previously, two studies were reported in literature on the structural phase transitions in LNN12. Based on the impedance spectroscopy techniques, Nobre and Lanfredi [8] reported that in the temperature range 300 to 1073 K, it undergoes four transitions. Recently, Mitra *et. al.* [22] also carried out temperature dependent dielectric measurements on $Li_xNa_{1-x}NbO_3$ upto 773 K and observed anomalies around 523 K and 623 K. They inferred that these observed anomalies correspond to structural phase transition from orthorhombic to cubic via tetragonal phase. But, this technique can not assign the correct crystal structure unlike diffraction methods. To the best of our knowledge, no further studies on the nature of high-temperature structural phase transitions have been reported.

In the present work, we have investigated the structural phase transitions in LNN12 in detail using high-temperature neutron and x-ray powder diffraction techniques and determined detailed the phase diagram of this important material. We observed interesting changes with appearance or disappearance of the super-lattice reflections in the powder diffraction patterns. This clearly reveals that LNN12 undergoes hitherto unreported structural phase transitions as a function of temperature. We infer that anomalies in dielectric measurements as reported in literature correspond to successive phase transitions from orthorhombic O2 (***Pmc2$_1$***) to O1 (***Cmcm***) to tetragonal (***P4/bmb***) to cubic (***Pm3m***), respectively. Subsequently, we suggest that the application of chemical pressure as a result of Li substitution in $NaNbO_3$ matrix favors the freezing of zone centre phonons over the zone boundary phonons that are observed in pure $NaNbO_3$.

## 2. Experimental

LNNx (x=0.06, 0.11, 0.12 and 0.13) samples are prepared by solid–state thermochemical reaction in the appropriate stoichiometric mixtures of $Na_2CO_3$, $Li_2CO_3$, and $Nb_2O_5$ at 1123 K for 6h in alumina crucibles. The calcined powders were sintered at 1423 K for 6h. The sintered pellets were crushed to fine powder and subsequently annealed at 823 K for 12h for removing the induced strain, if any, during the crushing process before using them for diffraction studies. X-ray diffraction studies were carried out using 18kW rotating Cu anode based powder diffractometer operating in the Bragg-Brentano focusing geometry with a curved crystal monochromator. Data were collected in the continuous scan mode at a scan speed of 1 degree per minute and step interval of 0.02 degree in the 2θ range of 20°–120°. The powder neutron diffraction data were recorded in the 2θ range of 7°–138°, using neutrons of wavelength of 1.2443 Å, on a medium resolution powder diffractometer at the Dhruva Reactor at Bhabha Atomic Research Centre. All the data collections were carried out during heating cycles of the sample for powder x-ray diffraction and cooling cycles for powder neutron diffraction measurements.

The structural refinements were performed using the Rietveld refinement program FULLPROF [23]. A Thompson-Cox-Hastings pseudo-Voigt with Axial divergence asymmetry function was used to model the peak profiles. The background was fitted using a sixth order polynomial. Except for the occupancy parameters of the atoms, which were fixed corresponding to the nominal composition, all other parameters, i.e., scale factor, zero displacement, isotropic profile parameters, lattice parameters, isotropic thermal parameters and positional coordinates, were refined. All the refinements have used the data over the full angular range. Although in the figures only a limited range is shown for clarity.

**Results and discussion**

**1. Evolution of powder x-ray and neutron diffraction data with composition and temperature**

Figure 1 depicts a portion of the powder x-ray and neutron diffraction patterns of LNNx with x= 0.06, 0.11, 0.12 and 0.14 at ambient condition. The diffraction patterns contain main peaks for a perovskite structure at around $2\theta \sim 32.5^0$ and $40.1^0$ (in x-ray data) and $26.5^0$, $32.0^0$ and $37^0$ (in neutron diffraction data), respectively. The peaks appeared around $2\theta \sim 36.2^0$, $38.5^0$ and $43.5^0$ in x-ray diffraction data and $2\theta \sim 29.2^0$, $30.9^0$ and $34.8^0$ are referred as superlattice reflection, which arise due to tilting of octahedral. It is evident from this figure that as Li concentration is increases above x= 0.11, an additional peak appears around 39.5 degree in x-ray diffraction patterns as marked with arrow. This additional reflection could be a signature for the presence of second phase, probably with rhombohedral symmetry in these compositions. However, due to medium instrumental resolution this feature is not visible clearly in neutron diffraction data.

To explore the possibilities of presence of rhombohedral phase in LNN12, we have refined the powder diffraction data using both orthorhombic (***Pmc2$_1$***) and rhombohedral (***R3c***) structure. The result of Rietveld refinements are shown in Figure 2. The fit between the observed and calculated profiles is satisfactory and all the reflections are accounted for using phase coexistence model. This confirms the presence of additional phase at ambient conditions. This suggests that additional peak could be treated as a signature for presence of rhombohedral phase.

Figure 3 shows the evolution of diffraction patterns as a function of temperature for LNN12. At the highest temperature (T= 1073 K), all the Bragg reflections present in the powder diffraction patterns could be indexed as main cubic perovskite reflections. On cooling, it undergoes a series of antiferrodistortive phase transitions. This structural phase transition driven by zone boundary instabilities result to multiplying the unit cell of the daughter phase. As a consequence of this, additional reflections appear in the powder diffraction patterns which are called as supelattice reflections. Based on the type of zone boundary instabilities these reflections appeared at different angles. Miller indices of these superlattice reflections based on an elementary perovskite cell present in powder diffraction gives information about the nature of the octahedral tilts in the structure. For example superlattice reflections with all-odd integered indices ("odd-odd-odd" i.e., "ooo" type in Glazer notation [24]) and two-odd and one-even integered indices result from anti-phase (− tilt) and in-phase (+ tilt) tilting of the adjacent oxygen octahedral. This is due to structural phase transitions driven by softening and freezing of the phonons at R (q= ½ ½ ½ ) and M (q= ½ ½ 0) points of the cubic Brillouin zone, respectively.

Now, we shall discuss the structures starting from the highest symmetry cubic phase that occurs at the highest temperature (T= 1073 K). All the Bragg reflections present in powder diffraction patterns could be indexed as main cubic perovskite reflections. As we lower the temperature, two new superlattice reflections appear at 2θ =29.5° and 35.0° (marked with arrow) in neutron diffraction patterns. These superlattice reflections are associated with zone boundary M point instability. Further, below 873 K, an additional set of superlattice reflections appear centered at 2θ =31.4° (marked with arrow), and associated with zone boundary R point instability. Further, lowering the temperature we found by enhancement of the intensity of superlattice reflection along with a dramatic change in Bragg's profile around 31.4°. Powder x-ray diffraction patterns also reveal appearance of superlattice reflection as observed in neutron diffraction study. For clarity, we have shown zoomed view at selected temperatures. It is noticed that the intensity of these reflections increases gradually with decrease of temperature and reflections are present at room temperature. The appearance and dramatic change in Bragg peak profiles clearly reveal occurrence of structural phase transition with temperature. It may be noted that after performing the temperature dependent neutron diffraction experiments, we found the signature for presence of the rhombohedral phase (similar to $LiNbO_3$ & about 2 %) even at highest temperatures and that the phase fraction does not change with temperature.

## 2. Phase stabilities in $Li_{0.12}Na_{0.88}NbO_3$ (LNN12) with temperature

As said earlier, all the perovskite materials have a stable cubic phase at highest temperature. Absence of superlattice reflections and splitting in main Braggs peaks above T> 925 K clearly suggests that LNN12 transforms to cubic phase. Thus, the powder x-ray and neutron diffraction patterns above 925 K are analyzed using cubic phase with the space group $Pm\bar{3}m$. The fit between the observed and calculated profiles is satisfactory (figure 4(a)) and indicating the correctness of model. As temperature is lowered below 900 K, we observed additional reflections (at 2θ =29.5° and 35.0° in neutron and ~ 36° in x-ray powder diffraction data) and splitting of main Braggs profiles suggesting the transition from cubic phase. Further lowering temperature below 873 K, an additional set of superlattice reflections appear centered at 2θ =31.4° (in neutron diffraction data) suggest the another transition. These superlattice reflections have odd-odd-even and odd-odd-odd type Miller indices and arise due to condensation of M and R zone boundary phonons and characteristics of tetragonal and orthorhombic phases respectively. Thus, for the temperature range (873 < T < 950 K) and (810 < T < 865 K) we refined the powder neutron diffraction data using tetragonal symmetry (space group: ***P4/mbm***) and orthorhombic symmetry (space group: ***Cmcm***), as observed in sodium niobate, respectively. The fit between the observed and calculated profiles is quite satisfactory. These structures index all the reflections including the weak superlattice reflections appeared in diffraction data and shown in Fig. 4(b) and (c).

Detailed Rietveld refinement of the powder diffraction data shows that diffraction patterns could be indexed using the orthorhombic structure (space group ***Cmcm***) up to 480K. The Rietveld refinements proceeded smoothly, revealing a monotonic decrease in lattice constant and cell volume with decreasing temperature. However, attempts to employ the same orthorhombic structural model in the refinements (Fig. 4 (d)) below 480 K, proved unsatisfactory, and a progressive worsening of the quality of the Rietveld fits with decreasing temperature is found. The most apparent signature of the subtle structural transformation that occurs at below 460 K is the inability of orthorhombic structure (space group ***Cmcm***) to account satisfactory for the peaks around 57.5 degree. For more clarity, it is shown in Fig. 4 (d) that the diffraction data at 573 K cannot be indexed with the orthorhombic phase. Extra broadening (splitting) of peaks suggests either lowering of the symmetry or coexistence of another phase.

### 3. Variation of structural parameters

Variation of lattice parameters with temperature obtained from the Rietveld refinements is plotted in Fig. 5. The [110], [1 -1 0] and [001] directions of the cubic phase correspond to the [100], [010] and [001] directions in the tetragonal (space group: ***P4/mbm***) and orthorhombic O2 (space group: ***Pmc2$_1$*** in different setting) phases, respectively. On the other hand, the lattice vectors of the orthorhombic O1 Phase (space group: ***Cmcm***) coincide with those of the original <100> cubic phase vectors. For the sake of easy comparison with the corresponding cell parameters of the various phases of LNN12, we have plotted equivalent elementary lattice parameters instead of lattice parameters corresponding to the different phases. It is evident from the figure that in the ferroelectric orthorhombic phase (O2 phase), the $a_p$ and $c_p$ parameters increase while the $b_p$ parameter decreases with increase in temperature. The discontinuous jump of lattice parameters at 633 K suggests probably a first order phase transition. Above 680 K, all the lattice parameters come close to each other and increase continuously with increase of temperature. As we further increase the temperature, $a_p$, $b_p$ and $c_p$ lattice parameter of the orthorhombic (***Cmcm***) phase increases and $a_p$ parameter approaches $b_p$ around 925 K. At this temperature, the sample undergoes an orthorhombic to tetragonal phase transition. Further as temperature increases, the $c_p$ lattice parameter decreases, and finally approaches to $a_p$ parameter, and the sample transforms into the cubic phase. The continuous change in the lattice parameters reveals that the successive phase transformations from orthorhombic O1 to high temperature tetragonal to cubic phase are not of a strong first order in nature.

The conductivity and permittivity of materials are known to depend on the crystal structure of the materials. Hence, any anomaly in these properties indicates structural rearrangement of the materials. Using impedance spectroscopy techniques, Nobre and Lanfredi [8] observed anomalies in permittivity data around 523 K, 721 K and 847 K. Based on these they reported that it undergoes four transitions from 300 to 1073 K. Recently, Mitra *et al* [22] also carried out temperature dependent dielectric measurements on $Li_xNa_{1-x}NbO_3$ upto 773 K and observed anomalies around 523 K, 623 K. They argued that these observed anomalies correspond to structural phase transition from orthorhombic to cubic via tetragonal phase. However, it is known that this technique cannot assign the correct crystal structure. As described in the above section, we observed distinct changes and appearance/ disappearance of the superlattice reflections in temperature dependence of powder diffraction patterns of LNN12. This clearly reveals that it undergoes a series of phase transitions as a function of temperature. By combining our present powder diffraction data and earlier work [8, 22], we have modified the phase diagram proposed by pervious workers [8, 22], which was mainly based on impedance spectroscopy studies. We now proceed to discuss the features of our phase diagram given in Fig. 5. At high temperatures, the structure of LNN12 is cubic. On lowering the temperature, it undergoes successively transition to tetragonal (T phase) and orthorhombic (O1 phase) phases as a result of the condensation of zone boundary M and R point instabilities, respectively. These two structural transitions are also observed in pure sodium niobate [18-19]. Further, on lowering the temperature, LNN12 undergoes ferroelectric orthorhombic O2 ($Pmc2_1$) phase which is linked with to freezing of zone centre phonon. This is in contrast to pure sodium niobate where its orthorhombic S, R and P phases with respect to orthorhombic ($T_1$) phase are linked to freezing of zone boundary phonons along the line T (lying between the M and R points) with (q= ½, ½, g) with g = 1/12, 1/6 and ¼ respectively. It is clear from this discussion, that application of chemical pressure as a result of Li substitution in $NaNbO_3$ matrix, favours the freezing of zone centre over the zone boundary phonon at ambient conditions.

## Conclusion

We investigated the phase stability of the technologically important material LNN12 using a combination of powder x-ray and neutron diffraction techniques in the temperature range 300- 1100 K. Distinct changes involving vanishing of the intensity of superlattice reflections, as observed in temperature dependence of powder diffraction patterns, clearly revealed that LNN12 undergoes a series of phase transitions as a function of temperature. Detailed Rietveld analyses of thermo-diffractograms show variety of structural phase transitions ranging from non-polar antiferrodistortive to ferroelectric in nature. Unambiguous experimental evidence is shown for phase coexistence of orthorhombic paraelectric (O1) phase (space group **Cmcm**) and orthorhombic ferroelectric (O2) phase (space group **$Pmc2_1$**) in the temperature range of 525 K to 675 K. The $a_p$ and $c_p$ lattice parameters of the ferroelectric orthorhombic phase (O2 phase), increase while the $b_p$ parameter decreases with increase in temperature. The discontinuous jump of lattice parameters at 633 K suggests a first order phase transition. Above 675 K, all the lattice parameters come close to each other and increase continuously with increase of temperature and $a_p$ parameter approaches $b_p$ around 925 K. At this temperature, the sample undergoes an orthorhombic to tetragonal phase transition. Further as temperature increases, the $c_p$ lattice parameter decreases, and finally approaches to $a_p$ parameter, and the sample transforms into the cubic phase. The continuous change in the lattice parameters reveals that the successive phase transformations from orthorhombic O1 to high temperature tetragonal to cubic phase are not of a strong first order type in nature. We infer that anomalies in dielectric measurements as reported in literature correspond to successive phase transitions from orthorhombic O2 (***Pmc2₁***) to O1 (***Cmcm***) to tetragonal (**P4/bmb**) to cubic (***Pm3m***), respectively. Subsequently, we suggest that the application of chemical pressure as a result of Li substitution in NaNbO$_3$ matrix favors the freezing of zone centre phonons over the zone boundary phonons that are observed in pure NaNbO$_3$.

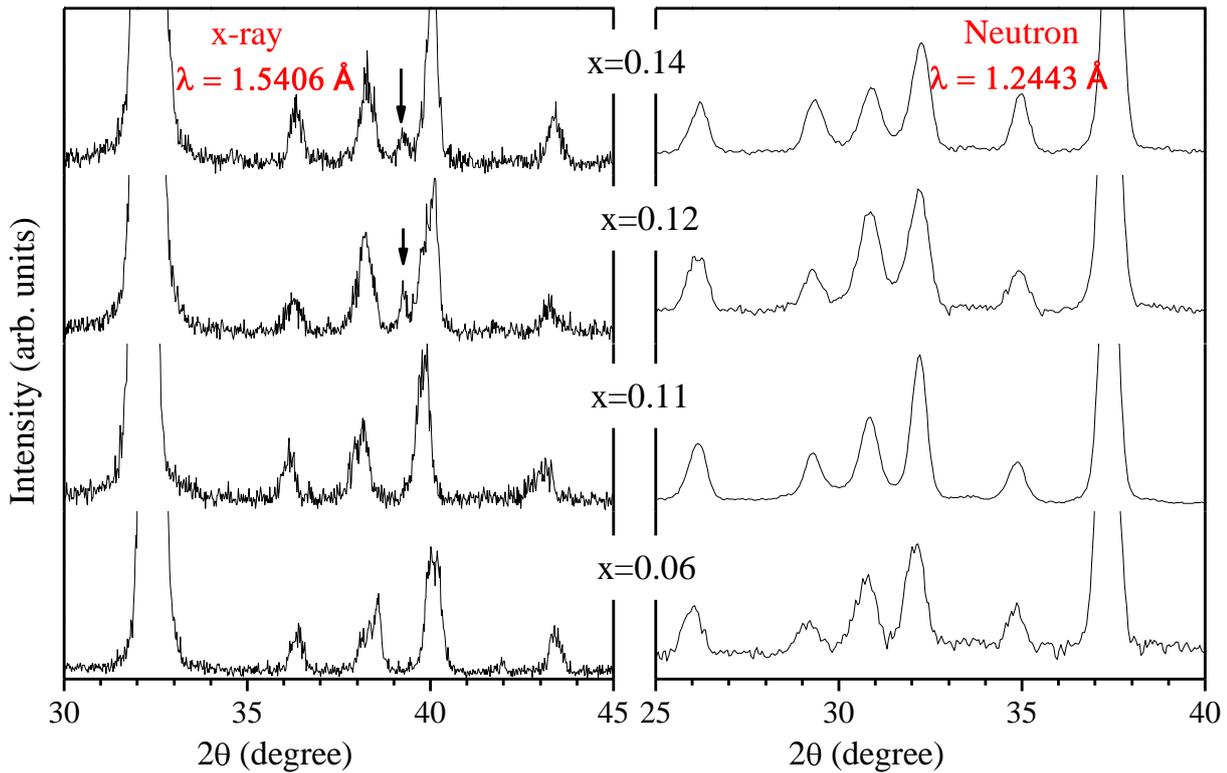

**Fig. 1** Evolution of a portion of powder x-ray and neutron diffraction patterns for LNNx with x=0.06, 0.11, 0.12 and 0.14 respectively at ambient temperature.

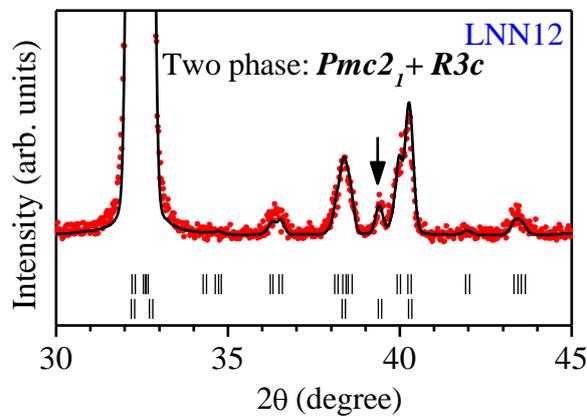

**Fig. 2** (color online) Observed (filled circle), calculated (continuous line) profiles obtained after the Rietveld refinement of x-ray diffraction pattern at 300 K of LNN12 using a mixture of ferroelectric orthorhombic ($Pmc2_1$) and rhombohedral ($R3c$) phases. The peak marked with an arrow is an additional peak. Upper and lower vertical tick marks correspond to the orthorhombic ($Pmc2_1$) and rhombohedral ($R3c$) phases, respectively.

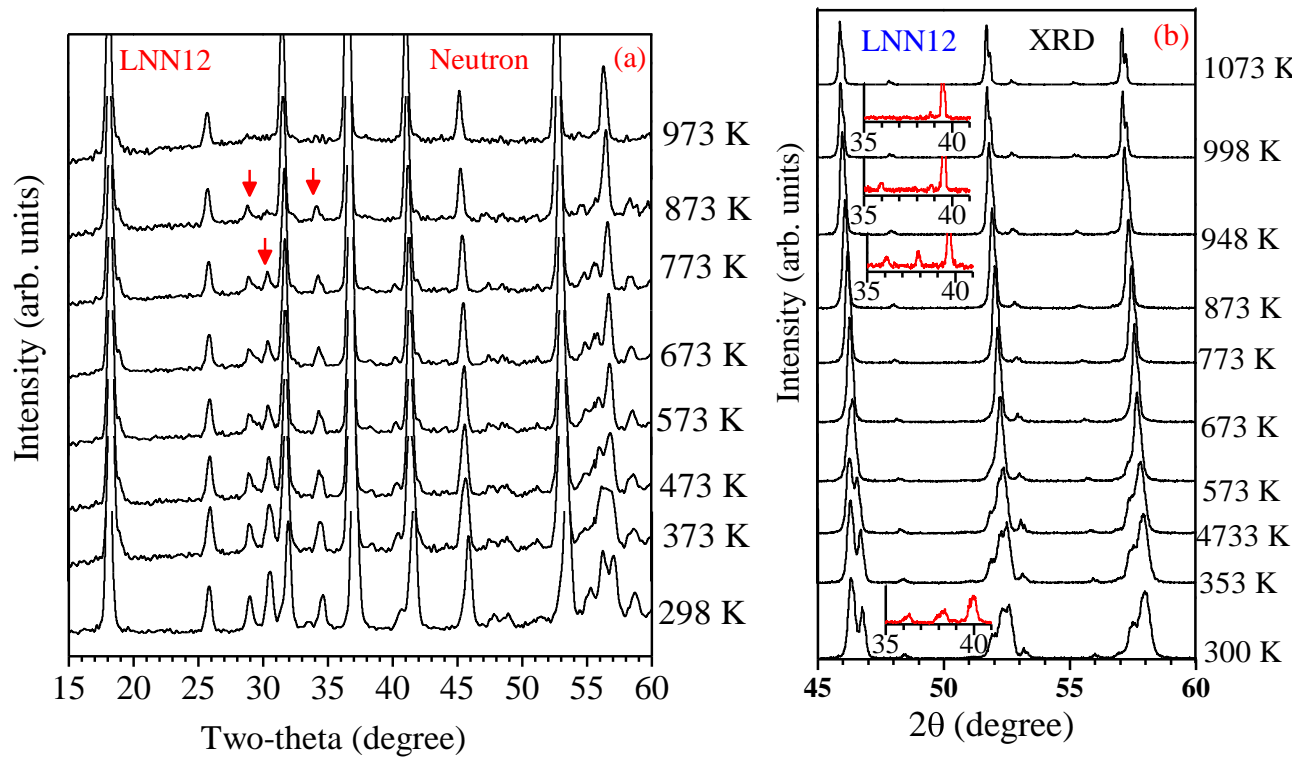

**Fig. 3** Evolution of a portion of powder (a) neutron and (b) x-ray diffraction patterns for LNN12 with temperature.

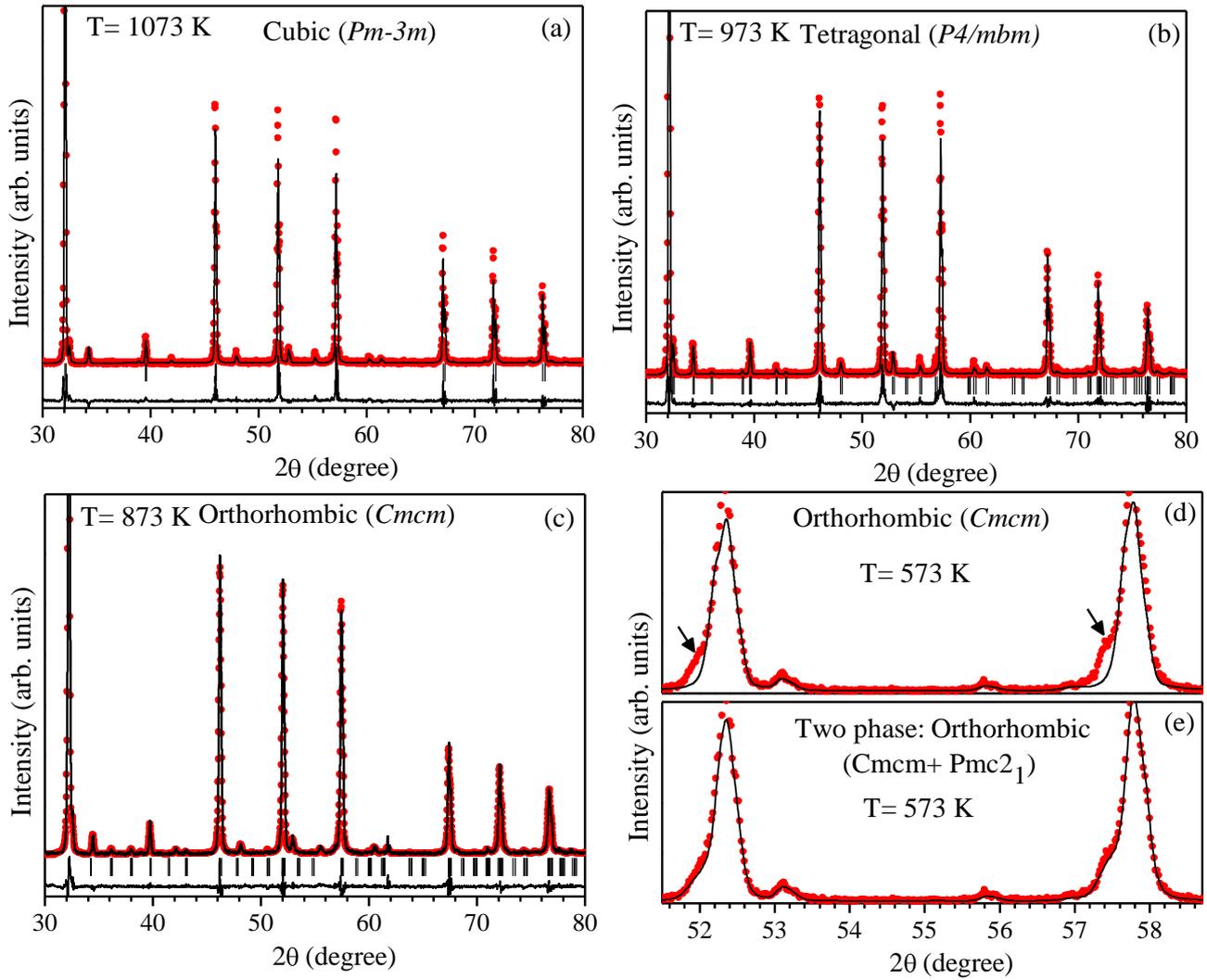

**Fig. 4** (color online) Observed (filled circle), calculated (continuous line) profiles obtained after the Rietveld refinement of x-ray diffraction pattern of LNN 12 using (a) Cubic (***Pm-3m***), (b) tetragonal (***P4/mbm***); (c &d) orthorhombic (***Cmcm***) and (e) two phase orthorhombic (***Cmcm*** + ***Pmc2_1***) at different temperatures, respectively. Peaks marked in (d) with arrows are additional peaks, which can be accounted only with two phase model in (e).

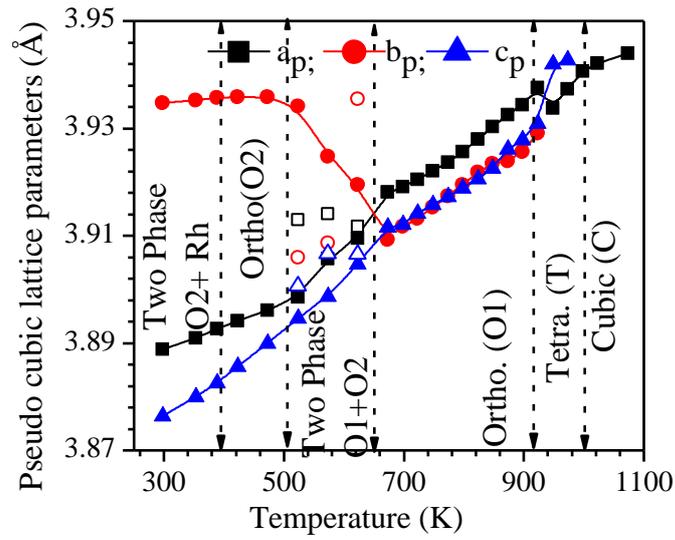

**Fig. 5** Variation of lattice parameters with temperature obtained from the Rietveld refinements of LNN12. In the coexistence region lattice parameter of minority phase is shown with open symbol. For clarity, the lattice parameters of ferroelectric rhombohedral phase are not plotted. At room temperature equivalent lattice parameters of rhombohedral phase are found to be $a_p$= 3.8691 (Å) and $c_p$= 3.9604 (Å) and phase fraction 28%, respectively.